\documentclass[aps,preprint,superscriptaddress]{revtex4-1}
\usepackage{graphicx}
\usepackage{amsmath,amssymb,bm}

\providecommand{\eqname}{eqn}
\providecommand{\sectionname}{Section}

\begin{document}

\title{Wetting and
cavitation pathways on nanodecorated surfaces} 

\author{M. Amabili}
\affiliation{Sapienza Universit\`a di Roma, Dipartimento di Ingegneria
Meccanica e Aerospaziale, 00184 Rome, Italy}
\author{E. Lisi}
\affiliation{Sapienza Universit\`a di Roma, Dipartimento di Ingegneria
Meccanica e Aerospaziale, 00184 Rome, Italy}
\author{A. Giacomello}
\affiliation{Sapienza Universit\`a di Roma, Dipartimento di Ingegneria
Meccanica e Aerospaziale, 00184 Rome, Italy}
\author{C.M. Casciola}
\affiliation{Sapienza Universit\`a di Roma, Dipartimento di Ingegneria
Meccanica e Aerospaziale, 00184 Rome, Italy}

\begin{abstract}
In this contribution we study wetting and nucleation of vapor bubbles
on nanodecorated surfaces via free energy molecular
dynamics simulations. The results shed light on the stability of
superhydrophobicity in submerged surfaces with nanoscale corrugations.
The re-entrant geometry of the cavities under investigation is capable
of sustaining a confined vapor phase  within the surface
roughness (Cassie state) both for hydrophobic and hydrophilic combinations of liquid
and solid. The atomistic system is of nanometric size; on this scale
thermally activated events can play an important role ultimately determining
the lifetime of the Cassie state. Such \emph{superhydrophobic} state can break down by 
full wetting of the texture at large pressures (Cassie-Wenzel transition) or by nucleating
a vapor bubble at negative pressures (cavitation).
Specialized \emph{rare event} techniques show that several pathways for wetting and cavitation
are possible, due to the complex surface geometry.  The related free
energy barriers are of the order of $100\;k_BT$ and vary with pressure.
The atomistic results are found to be in semi-quantitative accord with
macroscopic capillarity theory. However, the latter is not capable of
capturing the density fluctuations, which determine the destabilization
of the confined liquid phase at negative pressures (liquid spinodal).
\end{abstract}

\maketitle

\section{Introduction}

\begin{figure}[h]
        \centering
        \includegraphics[width=0.48\textwidth]{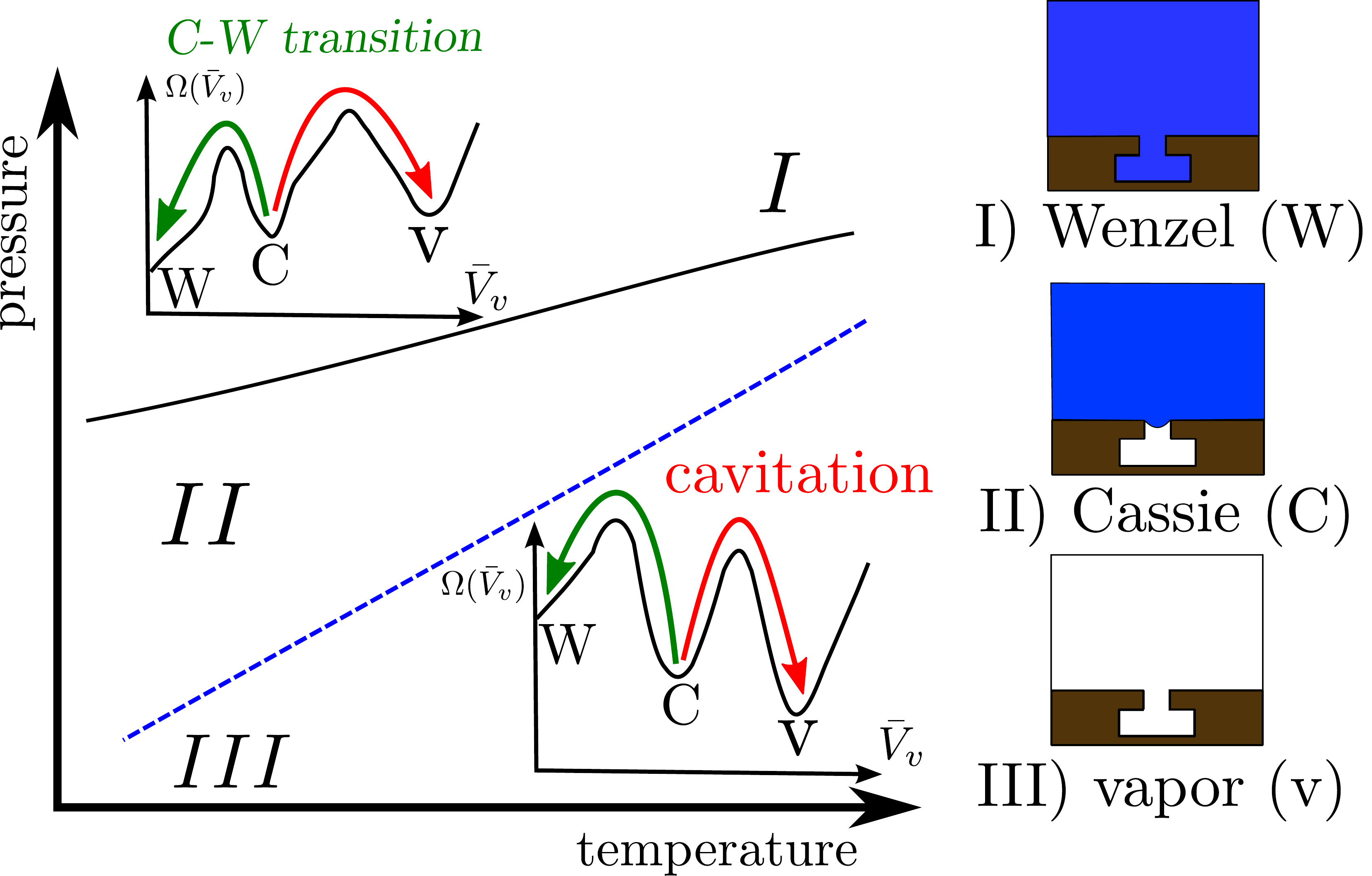}
        \caption{Sketch of the phase diagram for the Cassie,
								Wenzel, and vapor states. Typical free energy
								profiles (with the same notation as in the main text
								below) are shown in the region where the Wenzel state  (\emph{I})
								and the vapor state (\emph{III}) are thermodynamically
								stable, respectively. The Cassie state is thermodynamically
								stable in the intermediate region (\emph{II}).
                 The blue dashed line represents two-phase coexistence
								 for the bulk liquid and vapor phases along which
								 $P_l(T)-P_v(T)\equiv \Delta P(T)=0$. The black
                 solid line is the coexistence curve for the Cassie and
								 Wenzel states. 
								 The coexistence of the Cassie and 
								 vapor states coincides with the bulk two-phase
								 coexistence line at $\Delta P\to 0^-$, since for
							 	 arbitrarily small negative $\Delta P$ the pure vapor phase
							 	 is the thermodynamically stable one. The arrows
								 indicate the Cassie-Wenzel transition (green) and cavitation
						 		(red) through which superhydrophobicity is lost. }
        \label{fig:transizione}
\end{figure}

Confined fluids exhibit distinctive properties, which may significantly depart
from their bulk counterparts. A class of properties of
notable technological relevance is \emph{superhydrophobicity}, which proceeds
from the shifted liquid-vapor coexistence induced by confinement within
surface roughness (Fig.~\ref{fig:transizione}). Surfaces in the
superhydrophobic state are self-cleaning, ultra liquid repellent, and
may help reducing drag \cite{lafuma2003,zhang2008,nosonovsky2009,yan2011}.

However, the confined gaseous bubbles, which constitute the so-called Cassie state, may break down via two main
mechanisms \cite{Amabili2015} illustrated in Fig.~\ref{fig:transizione}:
intrusion of the liquid into the surface roughness, giving rise to the
so-called Wenzel state (Cassie-Wenzel transition), or formation of a critical vapor cavity (\emph{cavitation}) and its
subsequent dispersion in the liquid bulk. It is therefore of paramount
importance for technological applications to  determine quantitatively
the stability of the superhydrophobic Cassie state and
the kinetics of vapor loss.

A common strategy to stabilize the superhydrophobic state consists in decreasing
the characteristic size of surface roughness in order to increase the
capillary forces which can counteract the liquid pressure.
Present-day technologies \cite{checco2014a} are capable of
producing regular nanopatterns on arbitrarily large areas, which
have been exploited in the study of superhydrophobicity in nanoconfinement
 \cite{checco2014}. 
This ``shrinking'' strategy  is also exploited in hierarchical surfaces, which
combine a nanostructure for enhancing the bubble stability and a
microstructure for increasing the entrapped bubble volume
 \cite{verho2012}. In principle, it would be convenient to push the
miniaturization down to scales at which the
macroscopic models of capillarity show their limits \cite{Amabili2015}. This technological
challenge however calls for a more fundamental understanding of
superhydrophobicity at the nanoscale. In order to address these open
issues, here we use full-atom simulations and compare the results
with macroscopic capillarity theory.

The Cassie-Wenzel transition and cavitation through which
superhydrophobicity breaks down are, for most of the phase diagram illustrated
in Fig.~\ref{fig:transizione}, first order transitions, in which the
stable and metastable states are separated by free-energy barriers. On
the nanoscale, such barriers can be overcome by thermal fluctuations,
with the typical exponential dependence of the average time between
transitions on the free energy barrier \cite{eyring1935} (see \eqname~\eqref{eq:time}
below). For barriers larger than the thermal energy $k_B T$, this time 
becomes very long, giving rise to rapid but very infrequent
transitions, the so-called \emph{rare events}. Rare events are
challenging both from the experimental and from the simulative point of
view, because they require sampling of two very different timescales. 
In the present theoretical work, dedicated atomistic simulation
techniques -- the \emph{restrained molecular dynamics} (RMD) \cite{TAMD}
-- and continuum methods -- the \emph{continuum rare events method}
(CREaM) \cite{Giacomello2012} -- are adopted in order to
tackle rare events.

The aim of the present work is therefore one of investigating the
Cassie-Wenzel transition and cavitation on a nanodecorated surface.
The re-entrant structure with typical size of $5$~nm and the two
surface chemistries here considered -- hydrophilic and hydrophobic -- are similar
to those already explored in a recent communication \cite{Amabili2015}.
Such re-entrant geometries are exploited in order to realize the
Cassie state even with liquids having low surface tension
(omniphobicity) \cite{marmur2008,joly2009,savoy2012b,tuteja2008}.
Microscopic insights into the phenomena are
given by free-energy molecular dynamics simulations, which resolve
the atomistic structure of the fluids and of the solid. The validity of
the macroscopic capillarity theory -- in which the liquid-vapor
interface is sharp -- is assessed against them. We find that the two descriptions yield
surprisingly similar results; however, qualitative differences emerge close to the Wenzel
state due to the inherent compressibility of the atomistic model.
Multiple transition pathways are possible for the breakdown of superhydrophobicity
each characterized by different kinetics; the number of possible pathways
increases for hydrophilic surfaces.  Furthermore quantitative
differences in the free energy barriers are observed far from two-phase
coexistence. 

The paper is organized as follows. In the first section  the continuum rare
event method and restrained molecular dynamics  are briefly reviewed. In the
second section the results are discussed while the last section is left for
conclusions.


\section{Models and methods}

\subsection{The problem of rare events}
When a physical system is characterized by more than one (meta)stable
configuration, e.g., the Wenzel, Cassie, and vapor states in
Fig.~\ref{fig:transizione}, thermal fluctuations can 
drive the system between any two states.
When the transition occurs on a short timescale as compared to the
waiting time before the next thermally activated transition, we refer to this as a
\textit{rare event} \cite{bonella2012}.  
The minima of the system free energy (indicated in the following with
$\Omega$, see \eqname~\eqref{eq:grand-pot} below) correspond to the thermodynamically stable
configuration of the system (absolute minimum), or to the other
metastable states (local minima). 
The \textit{transition pathway} is defined as the collection of the
intermediate system configurations visited during the transition between
two states. As will be shown in the following, two states can be
connected by different transition pathways, each characterized by a
free-energy barrier $\Delta\Omega^\dag$, given by the free energy
difference between the free-energy minimum from which the transition starts 
and the maximum along the pathway (\emph{transition state}). The average time
$\tau$ needed for overcoming the free-energy barrier and thus
accomplishing a transition is given by \cite{eyring1935}:
\begin{equation}
	\tau \sim \exp{(\beta \Delta\Omega^\dag)}
\text{ ,}
\label{eq:time}
\end{equation}
where $\beta=1/(k_BT)$ is the inverse of the thermal energy with $T$ the
temperature of the system and $k_B$ the Boltzmann constant.  Thus if the
thermal energy $k_BT$ available to the system is less than the
free-energy barriers $k_BT < \Delta\Omega^\dag$, the transition
happens on a long timescale (rare). The barriers can be computed 
from the maximum of the free energy along a given transition pathway.

The free energy of a system can be expressed as a functional of some
relevant  quantity; e.g., in density functional theory it is the density $\rho(\mathbf{r})$ of the system
in the ordinary space $\mathbf{r}$ for which $\Omega \equiv
\Omega[\rho(\mathbf{r})]$.  In such a complex and  high dimensional
free-energy landscape computing the transition pathways and the related
free-energy barriers is a daunting task.  To alleviate this issue a reduced
set of  variables is often employed in order to characterize the
transition. The number and the particular expression of these variables
are dictated by the physics of the problem under investigation.
In this work we assume that the free energy depends on a single
variable, related to the filling of the cavity, see
Fig.~\ref{fig:sistemi}. For the atomistic case, this variable is
usually referred to as \emph{collective variable} (CV), reflecting the
fact that the atomistic degrees of freedom are mapped into a single
macroscopic observable. In section~\ref{sec:toy} we present a model
potential to illustrate the approximations
introduced by describing the system via a single variable.

\subsection{Continuum model and CREaM} \label{sec:CREaM}
\begin{figure}[h]
	\centering
	\includegraphics[width=0.48\textwidth]{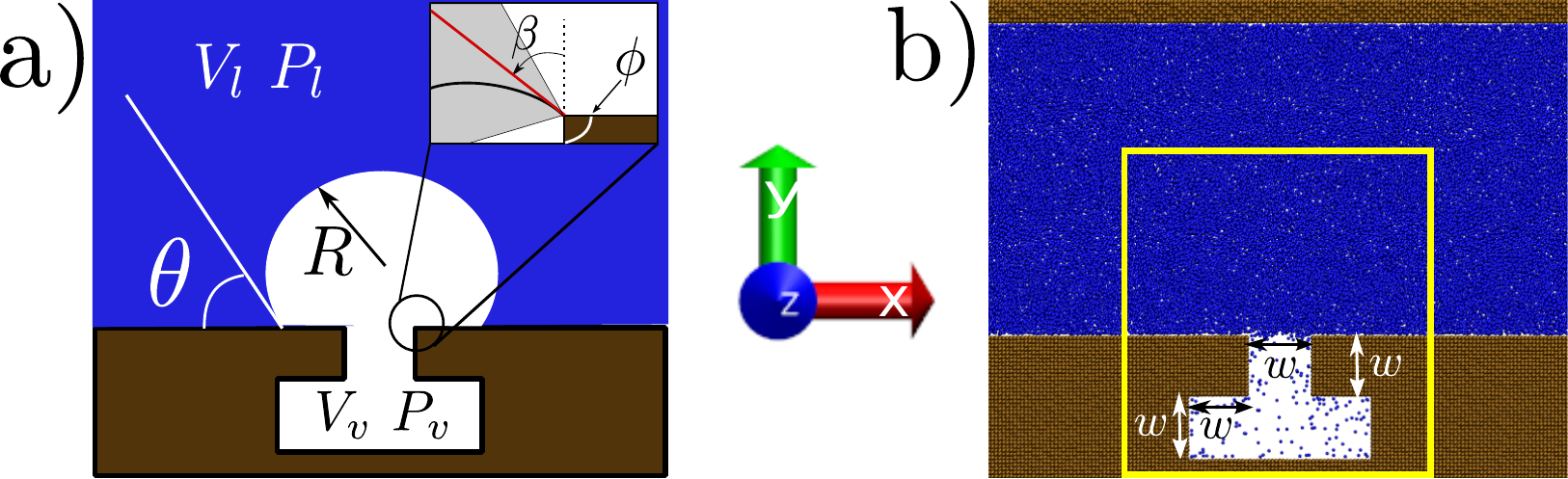}
	\caption{a) Sketch of the system used in CREaM calculations; the
		liquid phase is in blue, the vapor phase in white. The contact
		angle $\theta$, the radius of curvature $R\equiv R_1$ (in 2D,
		$R_2=\infty$), the corner angle $\phi$, and the angle $\beta$ used
		in the main text are defined. 
		b) Atomistic system used in RMD simulations; the fluid particles are in blue
		while the solid particles are in brown. The yellow box corresponds
		the control region used in the definition of the collective variable
		$N_{cav}$ (see \sectionname~\ref{sec:RMD}).}
    \label{fig:sistemi}
\end{figure}
Our capillary system is described via a sharp interface model, which
assumes that the properties of the liquid and vapor phases
are constant and equal to the bulk values up to the liquid-vapor
interface, where there is a sharp change in the properties. The
solid is fixed and forms a T-shaped cavity, see Fig.~\ref{fig:sistemi}. Using the
macroscopic capillarity theory, the thermodynamics of this
three-phase system can be described using the grand potential $\tilde \Omega$, which
depends on the chemical potential $\mu$, on the total volume $V$, and on the
temperature $T$:
\begin{equation}
\tilde{\Omega}(\mu,V,T)=-P_{l} V_{l}-P_{v} V_{v}+\gamma_{lv}A_{lv}+\gamma_{sv}A_{sv}+\gamma_{sl}A_{sl}
\text{ ,}
\label{eq:grand-pot}
\end{equation}
where $P_{l}$ and $P_{v}$ are the pressure of the liquid and vapor
phases, respectively, $V_{l}$ and $V_{v}$ their volume, $\gamma$ is the
surface tension of the corresponding liquid-vapor ($lv$), solid-liquid
($sl$) and solid-vapor ($sv$) interfaces, and $A$ their area. The
total volume of the system  available to the liquid and to the vapor
phases, $V=V_{l}+V_{v}$, is fixed.  Also  the total area of the solid surface
$A_{s}=A_{sv}+A_{sl}$ is constant.
These two constraints allow one to express the grand potential
parametrically in terms of the variables $V_v$, $A_{lv}$ and $A_{sv}$:
\begin{equation}
\Omega(\mu,V,T) \equiv \tilde{\Omega}-\Omega_\mathrm{ref}= \Delta P\mbox{ } V_v + \gamma_{lv} ( A_{lv} + A_{sv} \cos\theta_Y)
\text{ ,}
\label{eq:finale}
\end{equation}
where $\Omega$ is the excess grand potential, $\Omega_\mathrm{ref}=P_l V
+ \gamma_{sl} A_s$ is the reference grand potential computed in the
Wenzel state, $\Delta P\equiv P_l -P_v$, and $\theta_Y$ is the Young angle defined as:
$\cos\theta_Y\equiv (\gamma_{sv}-\gamma_{sl}) / \gamma_{lv}$.

By minimizing the grand potential functional in \eqname~\eqref{eq:finale}
w.r.t. $V_v$, $A_{lv}$, and $A_{sv}$, one finds the (meta)stable
configurations of the system. The conditions for stationarity are:
\begin{subequations}
\begin{align}
  &\Delta P = \gamma_{lv}\left(\frac{1}{R_1} + \frac{1}{R_2}\right) \text{ ,} \label{eq:Laplace} \\
  &\cos\theta = \frac{\gamma_{sv}-\gamma_{sl}}{\gamma_{lv}}  \text{ ,} \label{eq:Young}
\end{align}
\end{subequations}
where $R_1$ and $R_2$ are the principal radii of curvature of the liquid-vapor interface, 
taken to be positive if lying in the liquid domain and negative otherwise, and $\theta$ is the contact angle 
(see Fig.~\ref{fig:sistemi}). Equations~\eqref{eq:Laplace} and
\eqref{eq:Young} are the Laplace and Young equation, respectively. 
Equation~\eqref{eq:Young} holds at the liquid-vapor-solid contact line where the solid surface is smooth. 
At the corners of the T-shaped cavity, the normal to the solid surface jumps discontinuously. There 
the Young boundary condition is replaced by the Gibbs
criterion~\cite{Gibbs}, which requires that all the possible  
contact angles should be in the range $\theta_Y+\phi-\pi<\beta<\theta_Y$ 
(grey area in the inset of Fig.~\ref{fig:sistemi})  where $\theta_Y+\phi-\pi$ 
and $\theta_Y$ are the Young contact angles on the horizontal and
vertical surface, respectively.
Similarly, at the inner corners $\theta_Y<\beta<\theta_Y+\pi-\phi$ 
(see Fig.~\ref{fig:sistemi} for the definition of $\beta$ and $\phi$).  

The solutions of Laplace equation with Young or Gibbs boundary
conditions correspond to the stationary points of the functional in
\eqname~\eqref{eq:finale}, which can be minima (stable or metastable
states, e.g., Cassie or Wenzel), maxima, or saddle points (transition
states, e.g., a critical vapor bubble nucleating from the surface).
However, this procedure gives no information about the actual transition pathway(s) 
between two metastable states. In turn, the transition pathway selects the
free-energy barrier which is relevant for determining the kinetics of the
transition via \eqname~\eqref{eq:time}.

The Continuum Rare Events Method (CREaM) has been introduced to investigate
states beside the metastable ones and to construct the transition
pathways for the wetting or cavitation on arbitrary
geometries. The details of this formulation are discussed in
Refs.~\citenum{Giacomello2012} and \citenum{giacomello2013} and
here only briefly recapitulated. The basic idea consists in finding the constrained stationary points of
\eqname~\eqref{eq:finale} under the condition of a fixed volume of vapor
$V_v$, which is the progress variable used for the description of the
Cassie-Wenzel transition and cavitation. This procedure is carried out
using the method of the Lagrange multipliers. The constrained grand
potential is defined as $I=\Omega - \lambda(V_v-\bar V_v)$, where $\lambda$ is
the Lagrange multiplier and $\bar V_v$ is the chosen value for the volume of vapor.
Imposing the stationarity of the constrained functional $I$ yields \cite{Giacomello2012}: 
\begin{subequations}
\begin{align}
  & \Delta P - \lambda = \gamma_{lv}\left(\frac{1}{R_1} + \frac{1}{R_2}\right) \text{ ,} \label{eq:Laplace_mod}\\
  & \cos\theta = \frac{\gamma_{sv}-\gamma_{sl}}{\gamma_{lv}} \text{ ,} \label{eq:Young2} \\
  & V_v =\bar V_v \text{ ,} \label{eq:constr}
\end{align}
\label{eq:CREaM_sys}
\end{subequations}
where \eqname~\eqref{eq:Laplace_mod} is a modified Laplace equation in
which the Lagrange multiplier $\lambda$ plays the role of an extra
pressure term forcing the system to explore states beyond the metastable
ones. Equation~\eqref{eq:Young2} is the usual Young boundary condition
and \eqname~\eqref{eq:constr} enforces the volume constraint. 
Equations~\eqref{eq:CREaM_sys} can be solved numerically; here this task is
performed using the Surface Evolver~\cite{brakke1992}. 
Fixing the value of $\bar V_v$ and the appropriate boundary conditions
(\eqname~\eqref{eq:Young2} or the Gibbs criterion where needed) Surface
Evolver allows one to find the vapor configuration corresponding to a
minimum of \eqname~\eqref{eq:finale} under the fixed volume constraint.
From the triplet $(\bar V_v,A_{lv},A_{sv})$ computed by Surface Evolver it is
then possible to evaluate the free energy of the minimal configuration
via \eqname~\eqref{eq:finale} (see the ESI for details).
While \eqname s~\eqref{eq:CREaM_sys} in principle describe all the stationary
constrained states, the saddle points are not easy to detect via standard numerical
methods involving minimization, see, e.g., Ref.~\citenum{kusumaatmaja2015}. In practice only minima are accessible.
At a fixed $V_v$ multiple minima can be found, which correspond to different
configurations of the interfaces enclosing the same volume.
Repeating the numerical minimization procedure for various $\bar V_v$
between the Wenzel state ($\bar V_v = 0$) and large vapor volumes, which are
relevant to the cavitation regime, we are able to find the collection of
the constrained free-energy minima.
This information is used to construct (pieces of) possible transition pathways
between metastable states, which are characterized in terms of
the free-energy profile $\Omega(\mu,V,T;\bar V_v)$.

\subsection{Atomistic model and RMD} \label{sec:RMD}

The atomistic model used in molecular dynamics (MD) simulations consists of a Lennard-Jones (LJ) fluid,
defined by the pair interaction potential $V_{LJ}(r)= 4\epsilon[( {\sigma}/{r})^{12} - ({\sigma}/{r})^6 ]$, where
$\epsilon$ and $\sigma$ are the LJ energy and length scales,
respectively, and $r$ is the inter atomic distance between two fluid particles. The
solid walls (Fig.~\ref{fig:sistemi}b) are also made of LJ atoms; the lower wall is characterized
by a T-shaped nanocavity (Fig.~\ref{fig:sistemi}).
Solid and fluid atoms interact via a modified version of the LJ
potential: $V(r)= 4\epsilon[( {\sigma}/{r})^{12} - c ({\sigma}/{r})^6
]$. 
The coefficient $c$ multiplying the repulsive part of
$V(r)$ is used to tune the chemistry of the
surface.
The two LJ potentials completely define the interactions used in the atomistic system.  
The results coming from the MD simulations can be expressed in reduced units~\cite{Frenkelbook}, 
for example the reduced distance is defined as $r^{*}=r/\sigma$, 
the reduced pressure as $P^{*}=P \sigma^{3}/\epsilon$ and the reduced
temperature as $T^{*}=k_BT/\epsilon$.
From now on the superscript $*$, denoting dimensionless quantities, will be omitted. 
Two surface chemistries are considered in this work: a hydrophobic one
($c=0.6$) with a Young contact angle $\theta_Y=110^{\circ}$ and a
hydrophilic one ($c=0.8$) with $\theta_Y=55^{\circ}$.  Further details
on the choice of the coefficient $c$ and on the calculation of the
contact angle are found in Ref.~\citenum{Amabili2015}.

The atoms of the lower solid wall are kept fixed at their initial
positions forming an fcc lattice.  The upper solid wall, instead, is
used as a piston to control the pressure $P_l$ of the liquid phase. This is
achieved by applying to each particle belonging to the upper wall a
constant force in the $y$ direction (for details see
Ref.~\citenum{gentili2014} and the ESI).
A velocity Verlet scheme is used for the time evolution of the upper solid wall. 
The temperature of the liquid is kept fixed at $T=0.8$ via the
Nos\'e-Hoover chain thermostat \cite{martyna1992nose}.
The liquid temperature $T$, together with the liquid pressure $P_l$, sets
the thermodynamic conditions of the system.
The cavity dimensions are specified in Fig.~\ref{fig:sistemi}, with
characteristic length being $w \simeq 13$. The system extends for $7\,\sigma$ in
the $z$ direction.  Periodic boundary conditions are applied in the $x$
and $z$ direction. The position of the upper wall fluctuates along the
$y$ direction to enforce the required pressure, implying that the height
of the computational box changes in time.  In the simulation campaign we
explore a range of positive and negative liquid pressures, in the
interval $ -0.08 \leq P_l \leq 0.16$.

As anticipated before, the occurrence of rare events implies that MD trajectories are trapped in
the high probability regions of the phase space, the metastable states, and
transitions between these regions happen on a very long timescale as
compared to that accessible to MD simulations~\cite{bolhuis2002}.  A
convenient description of the system is given in terms of the collective variables
$\phi_i(\mathbf{r})$, which depend on the microscopic state of the system
$\mathbf{r}=\{\mathbf{r}_1,...,\mathbf{r}_N\}$.
For the system in Fig.~\ref{fig:sistemi}b the simplest choice is a single collective variable
counting the number of particles inside the T-shaped nano cavity (yellow
rectangle)~\cite{giacomello2012langmuir}.
This is a natural choice which can be directly related to the
volume $V_v$ of the vapor domain employed in the continuum approach in
\sectionname~\ref{sec:CREaM}.
Relevant information about the process can be extracted by computing the
probability that the observable $\phi(\mathbf{r})$ assumes a given value
$N_{cav}$, $p(P_l,T;\phi(\mathbf{r})=N_{cav})$, which depends also on the thermodynamic
conditions $P_l$ and $T$.  From $p(P_l,T;N_{cav})$ (where the dependence on
$\phi(\mathbf{r})$ is omitted) we define the Landau free energy:
\begin{equation}
	\Omega(P_l,T;N_{cav})=-k_{B}T\; \ln\, p(P_l,T;N_{cav}) \ .
\label{eq:Landau}
\end{equation} 
The microscopic expression above for the free energy can be compared
with the macroscopic grand potential profile $\Omega(\mu,V,T;N_{cav})$ found
via CREaM (see below).

Here $\Omega(P_l,T;N_{cav})$ is computed via the Restrained MD
method \cite{TAMD} (RMD), which amounts to adding  a biasing potential
of the form $V_{bias}(\mathbf{r})=k(\phi(\mathbf{r})-N_{cav})^{2}/2$  to the
physical one.  This harmonic-like potential, for suitable values of the spring constant $k$,
restrains the system close to $\phi(\mathbf{r})=N_{cav}$, forcing it to
explore also regions of the phase space with low probability which are
otherwise unaccessible to brute force simulations. 
Further details are found in reviews on rare event methods
\cite{ciccotti2011,bonella2012}; in brief, via RMD it is possible to
evaluate the gradient of $\Omega(P_l,T;N_{cav})$ according to:
\begin{equation}
	\frac{d \Omega(P_l,T;N_{cav})}{d N_{cav}} = \langle k(\phi(\mathbf{r})-N_{cav}) \rangle_{bias}
\text{ ,}
\label{eq:RMD}
\end{equation} 
where $\langle ... \rangle_{bias}$ represents the average computed over
the biased ensemble.  In practice the right hand side of
\eqname~\eqref{eq:RMD} is computed as the time average over a biased
MD simulation.  Each restrained simulation starts from an initial
configuration chosen in the Cassie basin.  The system is driven away
from the Cassie state by progressively changing the value of $N_{cav}$ until
the desired condition $\phi(\mathbf{r})=N_{cav}$ is reached.  After a standard
equilibration phase the statistics for the average in
\eqname~\eqref{eq:RMD} is collected. The spring constant is chosen to be $k=0.2$ which guarantees an accurate 
estimation of eqn~\eqref{eq:RMD}.
Summing up, in order to evaluate
the free-energy profile $\Omega(P_l,T;N_{cav})$, the free-energy gradient
(\eqname~\eqref{eq:RMD}) is computed on a set of equidistant 
points $N_{cav,i}$ via independent RMD simulations so that the full profile can be
reconstructed using a simple numerical integration:

\begin{equation}
	\Omega(P_l,T;N_{cav,M}) = \Omega_{0} + \sum_{i=0}^{M} \frac{d
	\Omega(P_l,T;N_{cav})}{d N_{cav}} \bigg|_{N_{cav}=N_{cav,i}} \, \Delta
	N_{cav}
\text{ ,}
\label{eq:ThermInt}
\end{equation}

where $\Omega_{0}$ is the free energy computed at  $N_{cav,0}$ 
and $\Delta N_{cav}=N_{cav,i+1}-N_{cav,i}=130$ is the fixed difference in $N_{cav}$ between 
successive RMD points.

The MD engine used in these simulations is  the open source code
LAMMPS~\cite{LAMMPS}. The biasing force is computed using the rare
events plugin PLUMED~\cite{PLUMED} which can be interfaced with LAMMPS.

\subsection{A simple two-dimensional example: the M{\"u}ller potential} \label{sec:toy}
\begin{figure*}
	\centering
	\includegraphics[width=1.\textwidth]{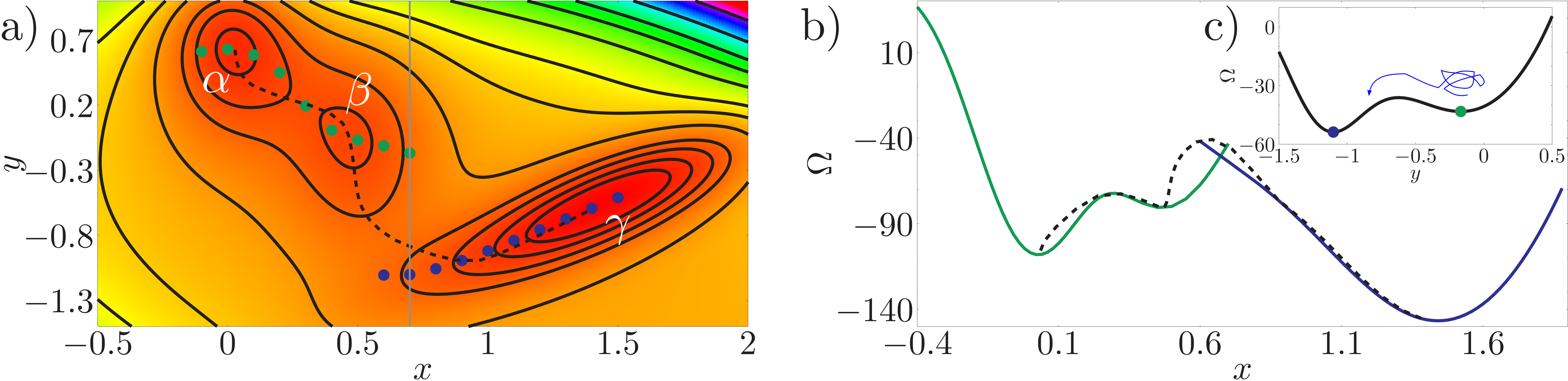}
	\caption{a) Color plot and isolines for the M{\"u}ller potential (arbitrary units; red
		corresponds to low values, while  violet to large ones). $\alpha$,
		$\beta$ and $\gamma$ are the minima of the potential. The blue and
		green points are the constrained minima along the line
		$x=\mathrm{const}$.  The black dashed curve is the actual transition
		pathway connecting the free-energy minima (taken from
		Ref.~\citenum{e2002}). b) Free-energy profile computed along the
		constrained minima (blue and green lines), and along the minimum
		energy pathway (dashed black line).  c) M{\"u}ller
		potential computed along the line $x=0.7$ (grey line in panel
		\emph{a}), showing two minima and the free-energy barrier separating
		them.   }
    \label{fig:rare_event}
\end{figure*}
The M{\"u}ller potential~\cite{muller1979} reported in
Fig.~\ref{fig:rare_event}a is a two dimensional potential
which is useful to illustrate rare events on rough free-energy
landscape, see, e.g.,
\citeauthor{weinan2007simplified}~\cite{weinan2007simplified}.
The M{\"u}ller potential is characterized by three (meta)stable states, labeled $\alpha$, $\beta$, and $\gamma$ in Fig.~\ref{fig:rare_event}a. 

Consider the case in which in order to describe the transition from
$\alpha$ to $\gamma$ only one variable is known (or can be observed),
say the $x$ variable in Fig.~\ref{fig:rare_event}; 
this is by construction an approximation since the M{\"u}ller
potential is two dimensional.
In this section we will try to clarify which are the approximations introduced with this reduced set of variables.
Applying CREaM in one variable to M{\"u}ller potential amounts to perform a constrained minimization on the line $x=const$. 
This procedure yields the set of minima at a fixed $x$, see, e.g., the blue and green points in
Fig.~\ref{fig:rare_event}a.
This set of points is then used to construct a candidate transition
pathway connecting the two minima, which is formed by piecewise smooth
branches roughly corresponding to the bottom of the valleys of the
potential. Finally, the value of the potential computed along the
pathways yields the free-energy profile reported in
Fig.~\ref{fig:rare_event}b with solid lines.

A similar result would be obtained by applying an RMD procedure to the same problem, 
provided that the MD trajectory remains confined to a ``valley''.
Indeed, the main difference between CREaM and RMD, even when the same
variable is used to describe the system, is the fact that atomistic
trajectories are affected by thermal motion. Thanks to thermal
fluctuations of the order of $ k_B T $ MD can escape from shallow minima.  For the M{\"u}ller
potential, depending on temperature, thermal fluctuations could overcome
the orthogonal barriers shown in Fig.~\ref{fig:rare_event}c, leading to
a more effective sampling of the phase space.  

Figures~\ref{fig:rare_event}a and \ref{fig:rare_event}b show that the cusps in
the free-energy profiles are a symptom of the presence of two neighboring
valleys. Around the cusp, the reduced description of the phase space via the
variable $x$ is insufficient. In RMD, due to the integration~\eqref{eq:ThermInt},
even smooth free-energy profiles can hide jumps between valleys (see, e.g.,
Fig.~\ref{fig:Salto_valli}) which become evident only by considering additional
observables (e.g., $y$ for the M{\"u}ller potential).

In this two dimensional example the exact transition pathway connecting
the metastable states $\alpha$ and $\gamma$ can be
computed~\cite{e2002} (black dashed line in
Fig.~\ref{fig:rare_event}a).  In Figure \ref{fig:rare_event}b the
free energy along the actual transition pathway is computed and
projected on the $x$ axis for comparison with the CREaM approximation.
As shown in Fig.~\ref{fig:rare_event}a,  CREaM or RMD solutions are
close to the exact transition pathway when the ``valleys'' are deep.
The two descriptions differ appreciably only near the transition state,
where the reduced description in terms of $x$ apparently breaks down.
However, the location of the metastable states and the barriers
separating them are similar.

Although the Cassie-Wenzel transition and cavitation are intrinsically
high-dimensional problems, a recent work~\cite{giacomello2015} has shown that
the scenario of a free-energy landscape with deep valleys illustrated in the
example above also applies to capillary problems similar to the present. In
other words the approximation in terms of a single collective variable is
generally viable, except in the vicinity of jumps between neighboring valleys.

\section{Results and discussion}

\begin{figure*}
        \centering
				\includegraphics[width=1.\textwidth]{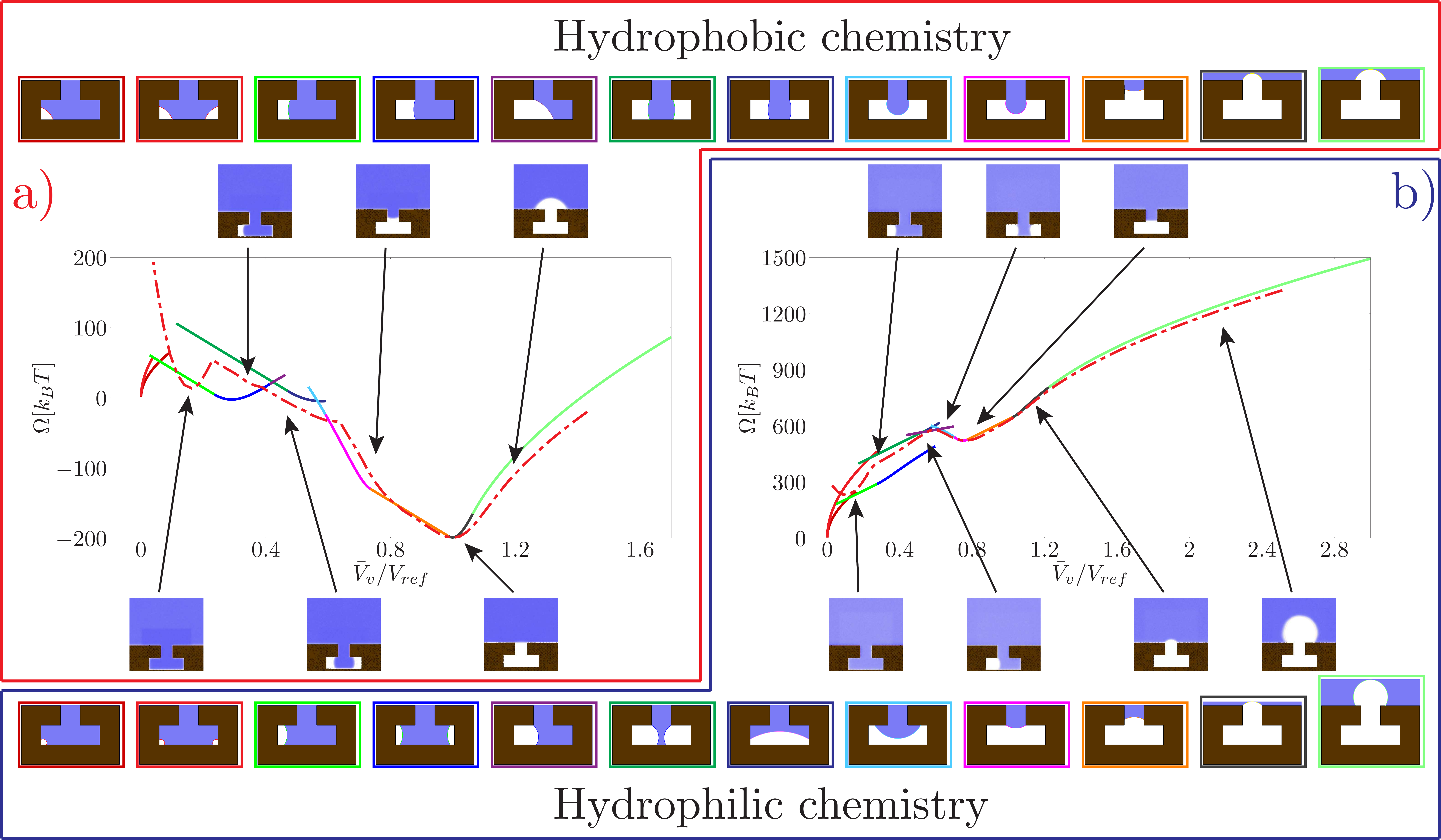}
				\caption{Free-energy profiles for the hydrophobic (a) and
					hydrophilic (b) chemistries.  The atomistic free energy is
					plotted with a red dash-dotted line.  The free energy
					computed via CREaM (\eqname~\eqref{eq:CREaM_sys}) is the solid
					line, with each color representing a different configuration
					of the liquid-vapor interface.  Such configurations are
					illustrated in the top and bottom strips for the hydrophobic
					and hydrophilic cases, respectively. The color of the
					rectangle enclosing each configuration corresponds to the
					branch of the same color in the CREaM free-energy profile. The
					insets show the atomistic average density field computed in
					RMD simulations, with the arrows indicating the corresponding 
				$\bar V_v$.}
        \label{fig:profili}
\end{figure*}

\subsection{Matching of continuum and atomistic parameters}

The continuum free energy defined in \eqname~\eqref{eq:finale} depends on few
thermodynamic parameters and material properties which need to be
specified in order to match the atomistic results:
$\Delta P$, $\gamma_{lv}$, and $\theta_Y$ in addition
to the geometrical dimensions of the system.
Equation~\eqref{eq:finale} describes a system at
constant chemical potential $\mu$, temperature $T$, and total volume $V$.
In particular the dependence of the free energy on $T$ and $\mu$ is via
the equation of state $\Delta P(\mu,T)$.  Thus, fixing the chemical
potential and the temperature is equivalent to fixing $\Delta P$. In the
atomistic simulations the system is characterized by a constant temperature $T$ and a constant
pressure $P_l$ in the bulk liquid. In addition, since the pressure $P_v$ of the
vapor phase depends primarily on $T$, fixing the temperature is
equivalent to fixing $P_v$. Hence $\Delta P=P_l-P_v$ is constant in
the MD simulations; in practice, $\Delta P$ is measured in simulations (for more details
see the ESI) and is set as input parameter for the macroscopic model. 
Two other physical parameters must be provided, i.e., the liquid-vapor surface
tension $\gamma_{lv}$ and the Young contact angle $\theta_Y$.
The surface tension is $\gamma_{lv}=0.57\pm0.02$ as estimated via
equilibrium simulations of liquid-vapor slabs (see ESI for details).
From the simulations $\gamma_{lv}$ is found to be independent of the
size of the liquid-vapor interface up to the investigated scale.
Finally the Young contact angle is computed following the same procedure of
Ref.~\citenum{Amabili2015}, which yields $\theta_Y=55^{\circ}$ and
$\theta_Y=110^{\circ}$ for the hydrophilic and the hydrophobic chemistry,
respectively.  Cavity dimensions are the same for both chemistries (see
Fig.~\ref{fig:sistemi}).

Having fixed the thermodynamic and material parameters, we need to find
a relation connecting the atomistic collective variable $N_{cav}$ and the
corresponding order parameter  $\bar V_v$ used in CREaM.
These two variables can be related using the so-called
sharp kink approximation according to which the bulk
properties of the liquid and vapor phases are extended up to the interface.
With this approximation, the total number of particles $N_{cav}$ in the
control volume is simply given by 
$N_{cav}=\rho_l V_l + \rho_v \bar V_v$,
where $\rho_l$ and $\rho_v$ are the bulk density of the liquid and of
the vapor phases, respectively. Considering that the total control
volume $V$ is fixed, $V=\bar V_v + V_l$, we can write  $N_{cav}=(\rho_v
-\rho_l)\bar V_v + const$. Here the constant is chosen such that $\bar V_v$
coincides for the atomistic and continuum cases at $ \Delta P = 0 $ (see
Fig.~\ref{fig:profili}).
Finally, in all the figures $\bar V_v$ is normalized with 
the volume $V_{ref}$ of the T-shaped micro structure, which
serves as a reference. 

\subsection{Free-energy profiles}

The free-energy profiles  at $\Delta P \simeq 0$ are reported in
Figs.~\ref{fig:profili}a and \ref{fig:profili}b for the hydrophobic and
hydrophilic case, respectively.  The CREaM solutions are plotted with
solid lines, each color encoding a different family of vapor domains.
The top and bottom rows of Fig.~\ref{fig:profili} show the corresponding
shapes of the vapor bubble for the hydrophobic and hydrophilic cases,
respectively. It is seen that at fixed $V_v$ several CREaM solutions are
possible; for visual clarity only those with lowest free energy are
shown in Fig.~\ref{fig:profili}. 
Both atomistic and continuum free-energy profiles show two minima for
hydrophilic and hydrophobic chemistries.  The minimum at $\bar V_v \simeq 0$
corresponds to the Wenzel state; while the minimum with $\bar V_v$ close to
unity is the Cassie state.  The shape of the meniscus in the Cassie
state is predicted by macroscopic capillarity theory: at $\Delta P \simeq
0$, the equilibrium condition, \eqname~\eqref{eq:Laplace}, renders an
infinite radius of curvature which corresponds to a flat interface. This
condition must hold together with the appropriate boundary
condition.  Thus an equilibrium Cassie state
can be obtained only when  the liquid-vapor interface is pinned at the
corner of the T-structure and the angle $\beta$  can take the value
$\beta=\pi/2$. According to Gibbs criterion explained in
\sectionname~\ref{sec:CREaM}, this  condition is attained at the outer
corner for the hydrophobic chemistry ($\bar V_v=V_{ref}$) and at the inner corner
for the hydrophilic one ($\bar V_v=0.75\,V_{ref}$). 

In CREaM profiles, the macroscopic Wenzel state is attained by
construction at $\bar V_v=0$ while the MD profiles have slightly
different values.  This discrepancy is due to the liquid density
depletion near the solid wall in the nano structure \cite{janecek2007}.
This depletion layer is not taken into account in the sharp interface
approximation used to relate $N_{cav}$ and $\bar V_v$.  Its thickness
depends on the chemistry of the surface and on the liquid pressure, with
the larger values corresponding to the hydrophobic solid and to low
pressures; this explains why the Wenzel state for the hydrophilic chemistry 
is closer to $\bar V_v =0$. 

In the following discussion, in order to make a quantitative comparison
between the microscopic and macroscopic description of the
Cassie-Wenzel transition and of cavitation, we divide the free-energy profile in
three regions
\begin{itemize}
	\item $\bar V_v \ge 0.7\, V_{ref}$ which is relevant to the cavitation regime;  
	\item $0.1 \, V_{ref}\le\bar  V_v \le 0.7\, V_{ref}$  which includes the configurations explored in the Cassie-Wenzel transition;
	\item $\bar V_v \le 0.1\, V_{ref}$ which corresponds to the Wenzel basin.
\end{itemize}
In the first region, the free-energy barriers and the critical volumes
are compared for atomistic and macroscopic models. Only the hydrophilic
chemistry is considered, since the critical bubbles for the hydrophobic
case are too large ( in the $x$-direction, see e.g. inset of Fig.~\ref{fig:profili}a for $V_v > 1$) 
for the definition of the collective variable given
in Fig.~\ref{fig:sistemi}.  In the second region, atomistic and
macroscopic Cassie-Wenzel  transition pathways are compared. Finally, in
the third region, the behavior of the two models near the Wenzel state
is analyzed.

\subsection{Cavitation}

\begin{figure*}
        \centering
				\includegraphics[width=1.\textwidth]{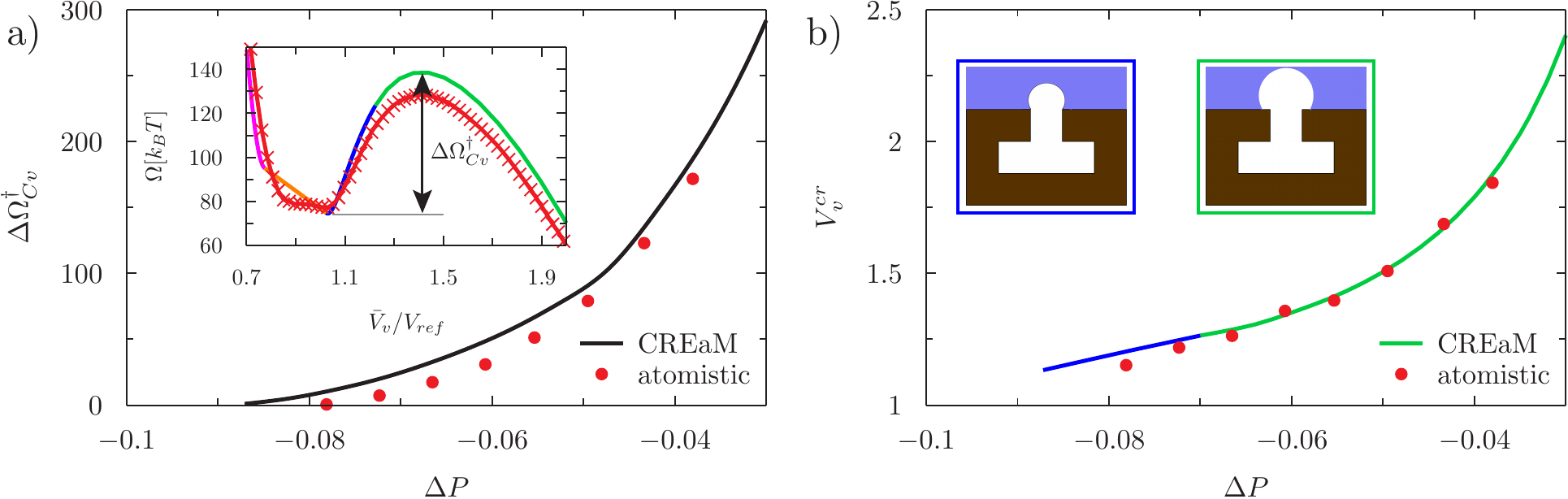}
				\caption{a) Atomistic (red symbols) and continuum (black line)
					free-energy barriers for cavitation as a function of $\Delta
					P$ computed for the hydrophilic surface. 
					The definition of the Cassie-vapor free-energy barrier $\Delta \Omega_{Cv}^\dag$ 
					is reported in
					the inset, showing a detail of the atomistic and continuum free-energy
					profiles for a representative negative pressure.  b) Critical
					volumes computed via atomistic (red symbols)	and continuum
					(solid lines) approaches as a	function of $\Delta P$. 
					Two different configurations of the critical bubble exist: a pinned bubble (blue line and 
					related inset) and a depinned one nucleating on the flat
					solid surface (green line and related inset).  }
        \label{fig:barriere}
\end{figure*}

At negative pressures $\Delta P<0$ the free-energy profiles show a maximum for $\bar
V_v \ge V_{ref}$ which corresponds to a critical cavitation bubble  of
volume $V_{v}^{cr}(\Delta P)$ (inset of Fig.~\ref{fig:barriere}a).  The
ensuing free-energy barrier $\Delta\Omega^\dag_{Cv}$ separates   the
Cassie state from the thermodynamically stable vapor state.  This
barrier, as stated before, dictates the kinetics of cavitation via
\eqname~\eqref{eq:time}.  Figure \ref{fig:barriere}a reports
$\Delta\Omega^\dag_{Cv}$ as a function of $\Delta P$ for the atomistic
simulations (red symbols) and for CREaM (black solid line). For the
simple cavitation pathways shown in Fig.~\ref{fig:profili}, the findings
of CREaM coincide with the classical nucleation theory
(CNT)~\cite{Volmer}.  The trend shows that the atomistic barrier is
always less than the macroscopic one.  These findings are in agreement
with previous simulation studies \cite{gonzalez2014,ten1998} which
predict that CNT overestimates the height of the barrier for the case of
homogeneous nucleation. The pressure at which $\Delta\Omega^\dag_{Cv}$
disappears is designated as \emph{spinodal} pressure for the Cassie-vapor transition,
$P_{sp}^{Cv}$; the atomistic value for $P_{sp}^{Cv}$ is less than the macroscopic counterpart.

The volume $V_{v}^{cr}$ of the  critical bubble is reported in
Fig.~\ref{fig:barriere}b as a function of $\Delta P$. For CREaM, the
critical volume (solid lines) is dictated by Laplace law
\eqname~\eqref{eq:Laplace}, which, in two-dimensions, gives $R_c=
\gamma_{lv}/\Delta P$ with $R_{c}$ the radius of curvature of the critical bubble. 
The critical volume $V_{v}^{cr}$ can be easily computed from $R_c$.
There are two possible configurations for the critical bubble. 
For largely negative pressures  the critical bubble is pinned at the outer corner of the 
T-shaped structure (blue line in Fig.~\ref{fig:barriere}b).
For moderately negative pressures, the critical bubble meets the
solid wall with the Young contact angle (green line in Fig.~\ref{fig:barriere}b).
The atomistic critical volume (red symbols in Fig.~\ref{fig:barriere}b)
are in fair agreement with the macroscopic ones at moderately negative
pressures for which the critical bubble is not pinned. In conclusion,
the Laplace equation predicts rather accurately $V_{v}^{cr}$ at the nano
scale and over a broad range of pressures.

\subsection{Pathways for the Cassie-Wenzel transition}
\label{sec:CW-transition}

\begin{figure}
        \centering
				\includegraphics[width=0.48\textwidth]{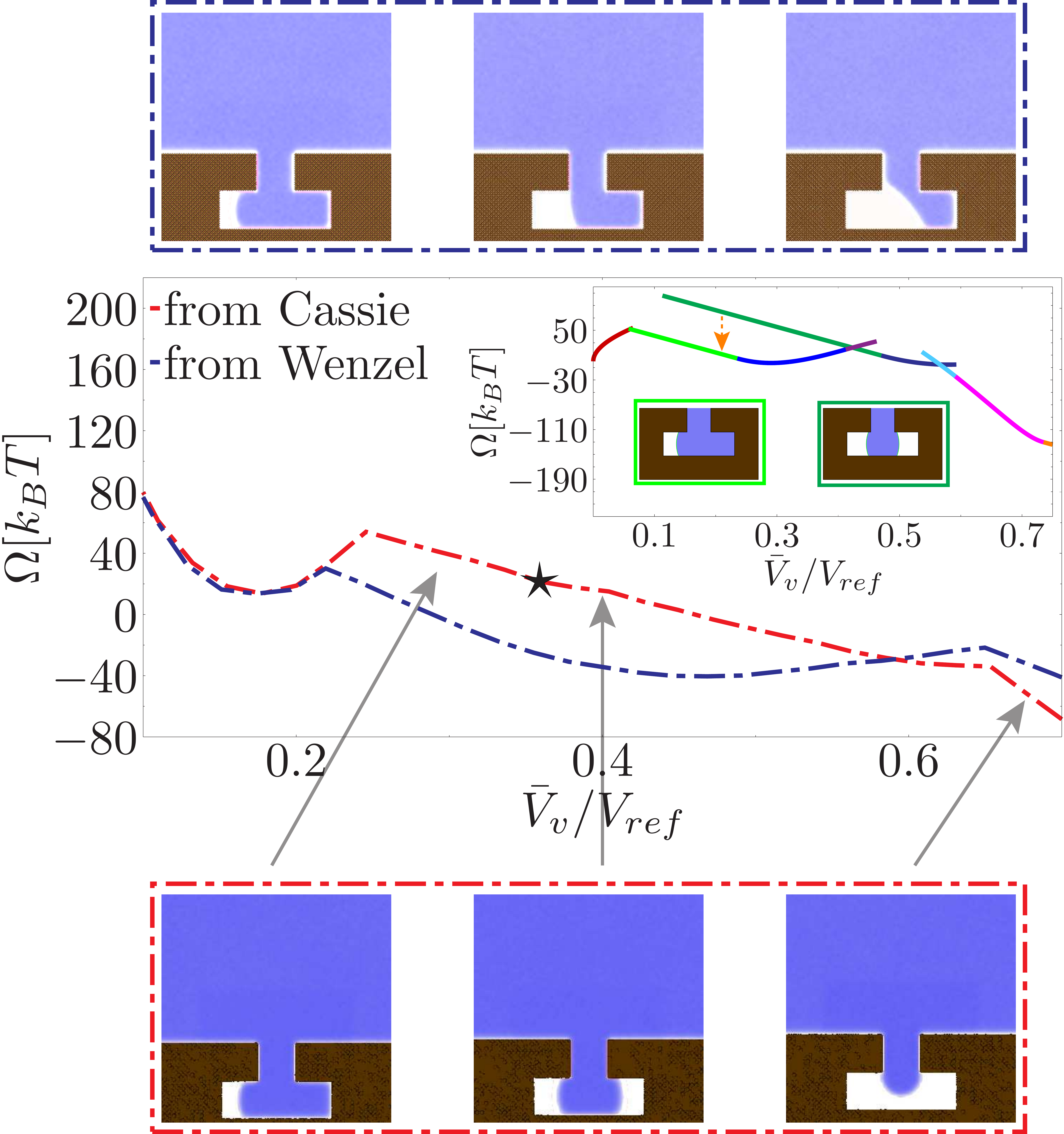}
				\caption{Detail of the atomistic (dash-dotted lines) and
					continuum (inset) free-energy profiles in the region of the
					Cassie-Wenzel transition for the hydrophobic chemistry. 
					The red (blue) curve represents RMD results
					started from atomistic configurations in the Cassie (Wenzel)
					state; the same color code applies for the average fluid
					density fields reported in the upper (Wenzel) and lower
					(Cassie) strips. The star in the main panel and the
					orange arrow in the inset identify the approximate location of
				  the jump between orthogonal valleys illustrated in the
					sketches above and explained in detail in the text.  }
        \label{fig:Salto_valli}
\end{figure}

Figure \ref{fig:Salto_valli} reports the free-energy profile and the
Cassie-Wenzel transition  pathway on a hydrophobic cavity for the
atomistic (red dash-dotted lines) and the macroscopic  approaches (inset). 
The initial conditions for these RMD simulations are in the Cassie state
as for the profiles in Fig.~\ref{fig:profili}.

Both RMD and CREaM free-energy profiles show cusps, which are signatures
of the transition between orthogonal shallow valleys as discussed in the
model potential of \sectionname~\ref{sec:toy}.  The first cusp, near the
Wenzel state, is discussed in detail in the next subsection. The second
cusp, at $\bar V_v \simeq 0.65 V_{ref}$ in the red RMD profile, corresponds to the
transition between an interface pinned at the inner corner (right
branch) and one with the liquid touching the cavity bottom (left branch,
see the lower strip in Fig.~\ref{fig:Salto_valli}).  

As the liquid touches the bottom, the atomistic transition pathway
switches between two different configurations: a symmetric one, with two
vapor bubbles in the arms of the T-structure and an asymmetric one, with
a vapor bubble in only one arm. In the macroscopic profile these two
states correspond to different solutions of the CREaM equations~\eqref{eq:CREaM_sys}. 
The occurrence  of this morphological
transition in RMD can be explained as a thermally activated jump between
two shallow valleys, as illustrated in Fig.~\ref{fig:rare_event}c.  In
other words, there is a free-energy barrier in the subspace orthogonal
to that spanned by the collective variable $N_{cav}$ which in RMD is
overcome by thermal fluctuations.  Clearly, the atomistic description
cannot follow each single branch separately because the orthogonal
free-energy barriers are of the order of $k_B T$.  On the opposite, the
macroscopic model is capable of following more branches but is unable to
jump between any two of them and to evaluate the orthogonal free-energy
barriers.  
The jump occurring in the atomistic simulations, identified by a star in
Fig.~\ref{fig:Salto_valli}, cannot be detected simply by looking at the
free-energy profile, which is smooth (no cusps). However, the jump
is easily revealed by examining the corresponding density fields along
the transition pathways. The observed abrupt change of configuration
corresponds to a jump between the two valleys identified by CREaM, shown
by the orange dashed arrow between the dark and light green profiles in
Fig.~\ref{fig:Salto_valli}. These two branches share, at least in CREaM, the same free-energy gradient. 
Based on the observed agreement between the atomistic and
macroscopic results, the same mean force is expected in
\eqname~\eqref{eq:RMD} irrespective of which of the two branches
is visited by the dynamics. In this way the morphological transition is
concealed by the thermodynamic integration used to reconstruct the RMD
profile (\eqname~\eqref{eq:ThermInt}).  The macroscopic expression for
the free energy, instead, is capable of distinguishing the (absolute,
with no undetermined integration constant) free energy of the two
branches revealing that the single bubble configuration is energetically favored.

A second atomistic pathway is reported in Fig.~\ref{fig:Salto_valli},
which is generated by choosing the initial condition for the RMD
simulations in the Wenzel basin (blue dash-dotted line).  It is seen
that the pathway selected by these initial conditions is different from
the one started from Cassie. In particular, the system dynamics cannot
explore the valley corresponding to two ``symmetric'' menisci because
this is at higher free energy by ca. $20\;k_BT$ (see the orange
arrow for CREaM calculation).
Thus the system is stuck in asymmetric configurations of the meniscus, which first pins at the lower
corner of the re-entrant mouth and eventually detaches from it (top
strip). This result confirms that the free-energy landscape is
extremely complex and suggests that for the Cassie-Wenzel transition and
for cavitation two different pathways can be followed. In order to
confirm this insight, more sophisticated techniques should be used, such
as the string method in collective variables \cite{maragliano2006}.

Similar arguments apply  also to the system with hydrophilic chemistry.
However in this case the number of alternatives valleys is very large
(see Fig.~\ref{fig:profili}b and ESI), making a detailed analysis like the one
reported for the hydrophobic case very cumbersome.

\subsection{Wenzel state}

\begin{figure*}
        \centering
				\includegraphics[width=1.\textwidth]{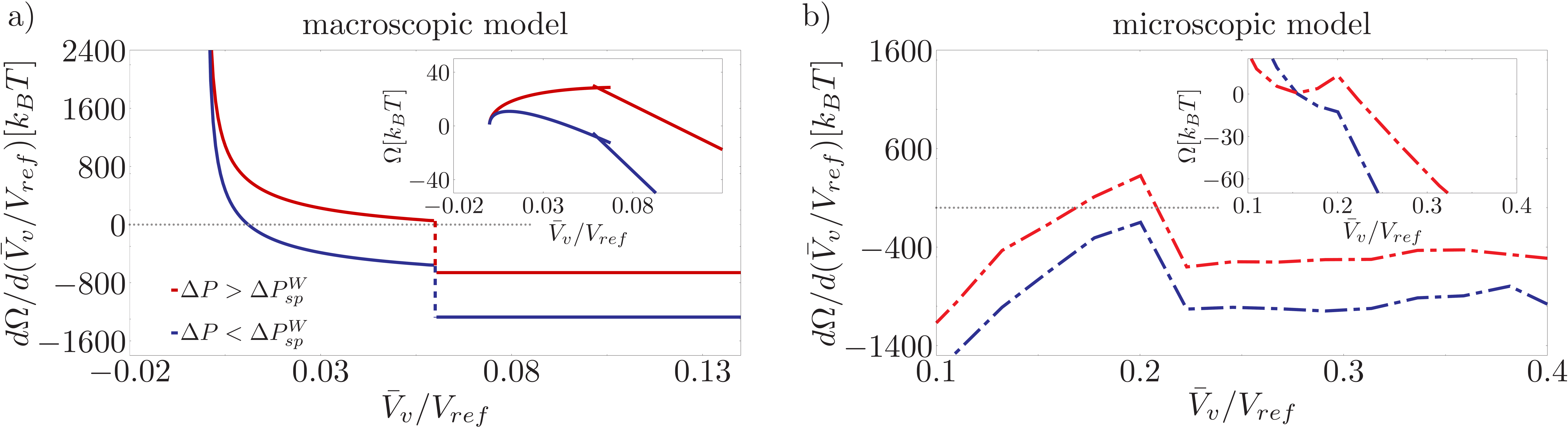}
				\caption{Derivatives of the free energy with respect to $\bar
					V_v$ (main panels) and free-energy profiles (insets) for the hydrophobic
					system obtained by CREaM calculations (a) and RMD simulations (b).
					Two pressures are reported: greater (red) and less (blue) than
				the liquid spinodal pressure $\Delta P_{sp}^W$.}
        \label{fig:spinodal}
\end{figure*}

Looking at the free-energy profiles in Fig.~\ref{fig:profili} it appears
that, for both chemistries, atomistic and macroscopic models have
qualitatively different behaviors close to the Wenzel basin: the concave 
atomistic free energy can be roughly described as a parabola with
an upward concavity in contrast with the macroscopic one which has
opposite curvature. Strictly speaking,
the parabolic approximation is  valid only around the Wenzel minimum at
$\Delta P=0$; for larger $\bar V_v$, deviations from parabolicity, the so-called fat tails \cite{patel2010}, 
cannot be excluded on the basis of the present computations.
The parabolic trend indicates that, close to the
Wenzel state, the probability distribution $p(N_{cav})$  for the
atomistic collective variable is Gaussian as per \eqname~\eqref{eq:Landau}.
This behavior is typical of liquids under confinement and accounts for
the fluid density fluctuations at the wall \cite{lum1999,patel2010}. 
These fluctuations can be related to the compressibility of
the (confined) liquid \cite{evans2015}.  
The upward concavity of the Wenzel
basin implies a positive compressibility, which is naturally
captured by the atomistic system.
For the macroscopic case, instead, the free-energy behavior is
completely different since the compressibility vanishes altogether by
the sharp-kink approximation ($\rho_l$ and $\rho_v$ are constant),
entailing a different trend. This can be made explicit by expanding close to
the Wenzel state the expression for the macroscopic free energy, which scales
as the liquid-vapor surface area $\Omega\propto \bar V_v^{2/3}$, which is quite different from the atomistic
scaling $\simeq  \bar V_v^2$.
Similar results are found by Remsing et al. \cite{remsing2015}, who
investigated the liquid/vapor transition between two flat hydrophobic
surfaces of nanometric extension. They also found that near the pure liquid
state the free energy is harmonic as opposed to the $\bar V_v^{2/3}$ trend
predicted by the (incompressible) classical nucleation theory. 
Another connection with the present results is the ``kink'' that these
authors find in free-energy profiles which could probably be interpreted in the
light of the simple model in Fig.~\ref{fig:rare_event}.

A direct consequence of the upward concavity of the Wenzel basin is
the existence of a liquid (or Wenzel) spinodal $\Delta P_{sp}^{W}$ which
has no counterpart in the (incompressible) macroscopic model, see the
insets in Fig.~\ref{fig:spinodal}b and in Fig.~\ref{fig:spinodal}a,
respectively.  This reflects the physical fact that
superhydrophobicity can be restored at sufficiently low pressures. On
the contrary,  the classical  nucleation theory fails to capture this
feature, predicting finite free
energy barriers for all pressures \cite{giacomello2013}.

In order to make these observation more quantitative, we
	report in Fig.~\ref{fig:spinodal}
 the free-energy profiles $\Omega(\bar
V_v/V_{ref})$ (insets) and their derivative $\mathrm d \Omega / \mathrm
d (\bar V_v/V_{ref})$ for pressures
greater (red) and less (blue) than the liquid spinodal.
For the macroscopic model, it is seen that the Wenzel state is attained
by construction at $\bar V_v = 0$, while the point at which the
free-energy derivative jumps from positive to negative values
corresponds to the free-energy maximum (Fig.~\ref{fig:spinodal}a).
Decreasing the pressure amounts to shifting the free-energy
derivative $\mathrm d \Omega / \mathrm d (\bar V_v/V_{ref}) \equiv\lambda$ by a
constant, see \eqname~\eqref{eq:CREaM_sys}.
Since, however, $\mathrm d \Omega / \mathrm d (\bar V_v/V_{ref})$ has a vertical
asymptote for $\bar V_v \rightarrow 0$, a maximum always exists in the continuum model.
This maximum is a regular point for large negative pressures, corresponding to a critical bubble nucleating 
in the corner, while it is a cusp for moderately negative pressures.
This cusp is generated by the presence of two valleys having different
slopes, cf. Fig.~\ref{fig:rare_event}.  

In the atomistic case, instead, the free-energy derivative changes
sign twice for $\Delta P \ge \Delta P_{sp}^{W}$. These two stationary
points are the Wenzel and the transition state, respectively, as shown
in Fig.~\ref{fig:spinodal}b. 
Upon decreasing pressures, the Wenzel state gradually shifts to larger
$\bar V_v$, see also Ref.~\citenum{patel2012}.
When $\Delta P \le \Delta P_{sp}^{W}$,  no
stationary point exists and the Wenzel state becomes unstable.  
	Because of the peculiar  shape of the Wenzel basin, 
in the atomistic case the maximum is always attained at the cusp where
the free-energy derivative has a discontinuity.

\section{Conclusions}
In the present work wetting and cavitation on nanostructured surfaces
have been studied via molecular dynamics and macroscopic
capillarity models. Rare events methods have been used in order to determine the wetting and cavitation
pathways and the related free energy barriers, which dictate the thermally
activated kinetics of the two phenomena. The systems considered here consist
of a re-entrant nano-cavity with hydrophobic and hydrophilic chemistry,
respectively. Given the re-entrant geometry, both chemistries allow for
the presence of a Cassie state.
We have found that the free energy landscape is characterized by many
``valleys'', indicating that many pathways are possible for wetting
and cavitation on nanostructured surfaces. These pathways and the
kinetics of the process strongly depend on the chemistry and on the
geometry of the surface, 
with the  hydrophilic chemistry showing the largest number of transition pathways. 


The present results allowed for a detailed quantitative comparison of
the atomistic and continuum models at the nanoscale. 
The major qualitative difference concerns the curvature of the free
energy profile close to the Wenzel state. For the atomistic model this
curvature is positive, accounting for density fluctuations of the
confined liquid \cite{lum1999}. The macroscopic
model, instead, due to the assumption of liquid incompressibility, does not
capture density fluctuations and features a negative curvature. This
discrepancy is reflected in the different (non-classical) pathways to wetting and
nucleation, which in turn lead to different estimates for the kinetics \cite{remsing2015}. Strictly
related to the compressibility is the presence of a liquid spinodal --
shifted by confinement as compared to the bulk one -- which is only
captured by the atomistic model. Quantitative differences emerge in the
free energy barriers connected with cavitation, with the macroscopic
model (classical nucleation theory) overestimating them. As
expected, the largest discrepancies are found away from two-phase
coexistence, where the size of the critical bubble becomes nanometric.

The complex free energy landscape connected with wetting and cavitation
on structured surfaces has required particular attention in interpreting
the results of the rare event atomistic simulations and continuum
calculations.  The approximation introduced by a reduced description of
the transition in terms of a single variable (the volume of the vapor
bubble) has been discussed in detail. This convenient choice, which is
normally used, e.g., in classical nucleation theory, is capable of
identifying most of the pathways, but fails when there is a
morphological transition \cite{giacomello2015}. 
Sufficient but not necessary symptoms of the failure of a reduced
description are the presence of cusps in the free energy profile. 
Furthermore, thermodynamic integration, which is often used in 
atomistic free-energy methods, may fail to capture significant free-energy jumps.
In order to overcome these limitations, a full description of the fluid density field would be required in order to
capture the details of the phenomenon and the exact free energy barriers.
These latter results call for a more profound
understanding of the coarse grained description of a
liquid.

\section*{Acknowledgements}
The research leading to these results has received funding from
the European Research Council under the European Union's
Seventh Framework Programme (FP7/2007-2013)/ERC Grant agreement n. [339446].  
We acknowledge PRACE for awarding us access to resource
FERMI based in Italy at Casalecchio di Reno.

\providecommand*{\mcitethebibliography}{\thebibliography}
\csname @ifundefined\endcsname{endmcitethebibliography}
{\let\endmcitethebibliography\endthebibliography}{}

\end{document}